\documentclass[preprint,5pt]{elsarticle}

\usepackage{amssymb,amsmath}
\usepackage{graphicx} 

\journal{Astronomy and Computing}

\begin{document}

\begin{frontmatter}

\title{GPU Accelerated Particle Visualization with Splotch}

\author{M.~Rivi\corref{cor1}}
\ead{Marzia.Rivi@astro.ox.ac.uk}
\address{Department of Physics, University of Oxford, OX1 3RH, United Kingdom}
      
\author{C.~Gheller}
\ead{cgheller@cscs.ch} 
\address{ETH-CSCS, via Trevano 131, 6900 Lugano, Switzerland}
  
\author{T.~Dykes, M.~Krokos}
\ead{timothy.dykes@port.ac.uk mel.krokos@port.ac.uk}
\address{School of Creative Technologies, University of Portsmouth, PO1 2DJ, United Kingdom}

\author{K.~Dolag}
\address{University Observatory Munich, Scheinerstrasse 1, D-81679 Munich, Germany}
\ead{kdolag@MPA-Garching.MPG.DE}

%\author{M.~Reinecke}
%\address{Max-Planck-Institut for Astrophysics, Karl-Schwarzschild Strasse 1, Garching, Munich, Germany}

\cortext[cor1]{Corresponding author}

\maketitle

\begin{abstract}
Splotch is a rendering algorithm for exploration and visual discovery in
particle-based datasets coming from astronomical observations or
numerical simulations. The strengths of the approach are production of
high quality imagery and support for very large-scale datasets through an
effective mix of the OpenMP and MPI parallel programming paradigms. This
article reports our experiences in re-designing Splotch for exploiting
emerging HPC architectures nowadays increasingly populated with GPUs. A
performance model is introduced to guide our re-factoring of Splotch.  A number of
parallelization issues are discussed, in particular relating to race
conditions and workload balancing, towards achieving optimal
performances. Our implementation was accomplished by using the CUDA
programming paradigm. Our strategy is founded on novel schemes achieving optimised data organisation and classification of particles. We deploy a reference cosmological simulation to present
performance results on acceleration gains and scalability. We finally
outline our vision for future work developments including possibilities
for further optimisations and exploitation of hybrid systems and emerging accelerators.
\end{abstract}


\end{frontmatter}

\section{Introduction}
\label{sec:intro}

The management and analysis of modern large-scale datasets generated by scientific experiments and numerical simulations can be very challenging due to continuously increasing sizes and complexity~\cite{DataTsunami}. Traditional data mining and analysis methods often rely on computationally complex
algorithms which can be very expensive if employed for 
large-scale datasets. Visual exploration and discovery can then represent invaluable tools, e.g.\ by providing scientists with prompt and intuitive insights enabling them to identify interesting characteristics and thus define regions of interest within which
to apply time-consuming methods. Additionally, they can be a very effective way in discovering and understanding correlations in data patterns, or in identifying unexpected behaviours, thus saving valuable resources, e.g.\ by terminating promptly ongoing numerical simulations producing unreliable results. Visualization tools can also provide effective means for communicating scientific results not only to researchers but also to members of the general public. The reader is referred to~\cite{DataAnalysisVisualization} for a recent discussion on visualization and data analysis challenges to address for next-generation large-scale datasets.

Astrophysics represents a prime example of a scientific field where visual data analysis and exploration is not just useful, but in some cases, mandatory. Often, in fact, no automated algorithmic solutions exist to identify and characterise features without requiring human assessment for correct interpretation. For this reason the astrophysical community has a long tradition in developing and deploying visual discovery tools for their datasets typically represented by images, complex surveys, data cubes or N-body simulations. Astrophysics thus represents an ideal discipline for exploiting High Performance Computing (HPC) devices, e.g.\ large multi-core and multi-node systems, providing all necessary resources for coping with large-scale datasets, e.g.\ computational power, large memory sizes, sufficient storage capacities and fast network speeds. A recent example is given by Hassan et al. (see related works and references in~\cite{2012ASPC..461...45H}), who focus on forthcoming generations of radio surveys (ASKAP~\cite{askap} and SKA~\cite{ska}), which are expected to produce enormous datasets. A further example, is given by Fraedrich et al.~\cite{Fraedrich:2009:TMV}, who
investigated scalability in visualizing large-scale particle-based cosmological 
simulations focusing on the Millennium run \cite{millennium}, and presented methods to reduce the associated limitations on PC architectures based on levels-of-detail.
Recently Kaheler et al.~\cite{2012arXiv1208.3206K} presented an algorithm specifically designed for N-body simulations. Their approach is based on a tetrahedral tessellation of the computational domain with mesh vertices defined by the simulation's dark matter particle positions and offers several GPU-assisted rendering solutions. This article includes an excellent review of methods for particle based visualization (see~\cite{2012arXiv1208.3206K} for details).

A number of popular, open-source software packages attempt to address the challenges associated with large-scale datasets and exploitation of HPC devices, e.g.\ VisIt~\cite{visit} and Paraview~\cite{paraview}, \cite{paraviewgrid}. Both are based on VTK~\cite{vtk} and support a fairly large variety of data types, file formats and visual discovery solutions. They can be used either as stand-alone tools locally or in a client-server configuration. Both tools support in-situ visualization~\cite{in-situ} allowing visualization procedures to be embedded in a simulation, thus generating images during a run (i.e.\ no data files are required) and enabling computational steering. A recent application of ParaView to large-scale cosmological simulations can be found in~\cite{2011ApJS..195...11W}. Neither of these packages provides customised tools for astrophysics, e.g.\ particle visualization capabilities are limited. In some cases the underlying computational performance and/or memory usage are not highly optimised, thus preventing effective deployment of these packages on modern, large-scale astrophysical datasets. Another VTK-based open-source software package focusing on astrophysics is VisIVO~\cite{visivo}. Although its main limitation is lack of interactivity, it has been recently released in the form of a science gateway offering a web-based, workflow-enabled framework seamlessly integrating large-scale, multi-dimensional datasets and applications for processing and visualization by exploiting distributed computing infrastructures~\cite{VisIVOGateway}. Advanced users are able to create, change, invoke, and monitor workflows while standard users are provided with customised web interfaces hiding all underlying technical aspects. Tipsy~\cite{tipsyurl}, Splash~\cite{splash}, GLnemo~\cite{glnemo} and Partiview~\cite{partiview} are further examples of software packages specifically designed for particle-based datasets. Applicability of these packages to large-scale datasets is limited as they operate in stand-alone mode without support for HPC resources.

This paper concentrates on {\it Splotch} \cite{2008NJPh...10l5006D} (see also the website \cite{splotch-web})
 which is a volume ray casting algorithm for effectively visualizing large-scale, particle-based numerical simulations. Very high-quality visualizations can be generated by Splotch for modern large-scale cosmological simulations, e.g.\ the Millennium trilogy~\cite{millennium}, the Horizon and MareNostrum runs~\cite{horizon} or the DEUS simulation~\cite{deus}. The underlying models in these simulations typically reproduce the evolution of a representative fraction of the universe by means of hundreds of billions of fluid elements (represented as particles), interacting through gravitational forces. The typical size of a time output (or {\it snapshot}) can range from several hundreds of GigaBytes~(GB) to tens of TeraBytes~(TB), recording ID, position and velocity of particles together with additional properties, e.g.\ local smoothing length, density and velocity. Although developed for numerical simulations, Splotch is being successfully employed also in other contexts, e.g.\ for visualizations of observed galaxy systems~\cite{m83-vis}.
The original Splotch algorithm has been optimised in terms of memory usage and exploitation of standard HPC architectures, e.g.\ multi-core processors and multi-node supercomputing systems by adopting the MPI paradigm~\cite{jin:high-performance} and OpenMP (see Sect.~\ref{sec:overview}).

Nowadays HPC systems are increasingly populated with GPUs employed not just as graphic accelerators but also as computational co-processors providing, on suitable classes of algorithms, outstanding performance with power efficiency significantly higher than standard CPUs. Supercomputers are thus increasingly equipped with several hundreds of GPUs that can overlap their computing capability with that of CPUs, minimising considerably overall times-to-solution of high-end scientific problems. This article discusses the issues we faced in designing and implementing a new version of Splotch using the CUDA paradigm~\cite{cudaurl} to fully exploit modern HPC infrastructures. The main issue was optimizing the rendering of variable radius particles, dependent both on their intrinsic size and on the point of view, posing several problems for the GPU computation in terms of race conditions and workload balancing. 

The same problem has already been faced by several authors, who proposed various solutions. In~\cite{liu2010}, two methods based on 
the adoption of CUDA optimized atomic operations are used to implement on the GPU a full rendering pipeline, called {\it FreePipe}. 
This approach provides good performance on relatively large datasets of the order of few million elements.
%when small triangles of homogeneous size have to be drawn. 
For much larger data, as that Splotch addresses, atomic operations 
can lead to performance penalties. Furthermore, our data encompasses particles of any radius (from very small, to the size of the full image),
for which FreePipe is not optimal. In~\cite{laine2011}, a more sophisticated and comprehensive solution
is proposed, based on the implementation of several rendering steps, acting at different resolutions and allowing performing the whole
data processing on the accelerator. 
Such an approach however, requires a complete refactoring of the code and introduces additional memory consuming algorithmic components that may impact overall performance for large data sizes. 
The tests presented in their paper, in fact, 
deal with only at most few million points, a small number 
compared to our target data sizes, spanning from hundred million to billion particles.
As a result we propose a novel approach, based on a concurrent use of the GPU and the CPU.
This followed a
preliminary investigation, as reported in~\cite{jin:high-performance}, that 
had to be redesigned from the ground up to significantly improve performance. A number of further 
optimised solutions for rendering particles have been implemented, based on a novel data classification and organization strategy. Furthermore, the Thrust library~\cite{thrusturl} was adopted for an optimal implementation of specific kernels requiring sorting or reduce-by-key operations.
This improved performance is achieved without affecting the linear scalability on the number of particles of the original Splotch. 

The paper is organised as follows. 
Section~\ref{sec:overview} is a short description of Splotch. The CUDA paradigm and a performance model that guided our re-designing of Splotch are discussed in Sect.~\ref{sec:gpu-code}. Our implementation is presented in Sect.~\ref{sec:implementation} in which we describe our strategy on how to classify particles for rendering. Section~\ref{sec:results} presents our reference datasets for benchmarking and discusses performance results including scalability related to sizes of datasets and smoothing radius. Finally Sect.~\ref{sec:conclusions} presents conclusions and pointers to future developments.

\section{Splotch Overview}
\label{sec:overview}

Splotch generates high-quality images through a customised volume ray casting approach, described in detail in \cite{2008NJPh...10l5006D}.  
The implementation of Splotch is self-contained with no dependencies from external libraries (apart of course from those needed for parallelism and to support specific file formats, e.g.\ HDF5 \cite{hdf5}). The code is pure C++ and can be compiled through suitable makefile scripts.
The main stages of Splotch are summarised below:
\begin{itemize}
\item
{\it Data Loading} - Various readers are available supporting numerous file formats. At least three scalars are required representing particle coordinates in a user-defined coordinate system.
\item
{\it Processing and Rasterization} - Normalization together with other necessary 
calculations (e.g.\ for logarithms of processed quantities) are performed. Particle
coordinates and other geometric quantities (such as smoothing lengths - see below) are
roto-translated and projected according to camera settings.
{\it Active particles},  contributing to the rendering, are identified and assigned with RGB color components. The remaining
{\it inactive particles}, that, depending on the camera position, lie completely outside the scene, are not subject of any further processing.
For simplicity this stage will henceforth be referred to as {\it Rasterization}.
\item
{\it Rendering} - The contributions of individual particles to the final rendering are calculated by solving the radiative transfer equation  \cite{1991par..book.....S} along lines of sight originating from each pixel:
\begin{equation}\label{rad}
\frac{d{\bf I}{(\bf x})}{dr}=({\bf E}_p-{\bf A}_p{\bf I}({\bf x}))\rho_p({\bf x}),
\end{equation}
where ${\bf I}({\bf x})$ represents radiation intensity at position ${\bf x}$, $r$ corresponds to a coordinate along the line of sight,  ${\bf E}_p$ and ${\bf A}_p$ are the coefficients of radiation emission and absorption of particle $p$ respectively and $\rho_{0,p}$ is a physical quantity (e.g.\ mass density or temperature) transported by particle $p$ according to a Gaussian distribution:
\begin{equation}\label{kernel}
\rho_p({\bf x}) = \rho_{0,p}\exp(-{||{\bf x}-{\bf x}_p||}^2/\sigma_p^2),
\end{equation}
where ${\bf x}_p$ denotes particle coordinates. This distribution is clipped to zero at a given distance $\chi\cdot\sigma_p$, 
where $\chi$ is a suitably-defined multiplicative factor and $\sigma_p$ is the particle smoothing length. Any rays passing at distances larger than $\chi\cdot\sigma_p$ are unaffected by $\rho_p$. 
Assuming ${\bf E}_p = {\bf A}_p$ the solution of equation \eqref{rad} does not depend on the particles integration 
order along the line of sight, thus strongly simplifying the design of a parallel 
version of the algorithm.
% that splits the data among processors each calculating an independent partial image. 
This approach has proved to be
%This condition cannot be adopted in general cases. However, after extensive testing on a number of different datasets,  it proved to be 
effective for typical astrophysical data, due to the fact that visualised matter is either diffused and optically thin 
(i.e. almost transparent, as for the intergalactic medium) or extremely bright and compact (as for stars or small galaxies).
%In these opposite regimes the ${\bf E}_p = {\bf A}_p$ assumptions gives no artifacts on the resulting images. 
%However, there are cases where this condition leads to negligible errors (as for gas clouds in spiral galaxies), and
%for which it can still be adopted, and other cases where it gives wrong images (as in the case of dust clouds). 
%In this last cases, independent emission and absorption coefficients can be adopted. This option is available
%only in the serial version of the code. 
The two coefficients 
%can vary between particles and 
are typically chosen as a function of a characteristic particle property (e.g.\ temperature or density).
Equation \eqref{rad} is solved for each color component (R, G and B) separately. These components can be defined based on relevant physical quantities, e.g.\ for velocity $\bf v$$=(v_x, v_y, v_z)$ we assign $\rho_p^{R}=v_x$, $\rho_p^{G}=v_y$ and $\rho_p^{B}=v_z$. Conversely, a scalar quantity, such as mass density or temperature, can be mapped to RGB components via look-up tables (or palettes). 
\end{itemize}
Large-scale datasets are supported by exploiting HPC architectures by means of an effective mix of OpenMP / MPI parallel programming paradigms. The MPI implementation~\cite{jin:high-performance}  simply distributes chunks of particles among different processors, each performing a serial computation and producing a partial rendering, the root processor composing final renderings. The OpenMP implementation performs {\it Rasterization} exploiting multiple threads each working on different chunks of particles. Regarding {\it Rendering}, images are subdivided into tiles (e.g.\ $100 \times 100$ pixels) which are processed by OpenMP threads on a ``first-come, first-serve" basis. To determine the particles to be considered for individual tiles the entire particle array is processed prior to rendering, so that for each tile a list of particle indices is generated.

\section{GPU Considerations}
\label{sec:gpu-code}

Splotch has been realised on modern GPUs by exploiting the CUDA programming paradigm. This section reviews the fundamentals of CUDA programming and introduces details of a performance model we developed to guide the code refactoring discussed in Sect.~\ref{sec:implementation}.  

\subsection{CUDA Paradigm} 
\label{sec:cuda}

The CUDA programming model \cite{cudaurl} by NVIDIA currently represents the standard ``de facto'' for GPU programming. The underlying GPUs are closely mapped leading to optimal performance. The obvious drawback of limited portability is somewhat mitigated by the popularity of NVIDIA GPUs. Access to highly parallelised modern GPU architectures is offered through a simplified C/C++ or Fortran API supporting joint CPU/GPU execution. Serial procedures are performed by the CPU (host), while those which exhibit rich amount of data parallelism are performed by the GPU (device) realised as CUDA kernels. The CPU and GPU have their own memory space so any data transfers must utilise the PCI Express bus (for simplicity hereafter referred to as PCI-E).

The launch of CUDA kernels is asynchronous allowing the host to execute code instructions while the device is computing. If the host requires a kernel execution to be completed then it is necessary to call a device synchronization function. A kernel is instantiated as a grid of lightweight parallel threads organised into {\em blocks} of the same size. A thread represents an independent work element and maps to a hardware core or streaming processor (SP). A block is a 1D, 2D or 3D set of concurrently executing threads that can cooperate among themselves via barrier synchronization and fast shared memory. This is possible because threads belonging to the same block are executed on the same streaming multiprocessor (SM). However synchronization is not possible between blocks and there is no guaranteed order for block execution. Moreover the limited amount of shared memory limits the number of  blocks that can reside in a particular SM simultaneously.  

Once a block is assigned to an SM, it is partitioned into 32-thread units called {\em warps}. All threads belonging to a particular warp are scheduled to execute the same instruction (Single-Instruction, Multiple-Thread). To optimise overall performance, programmers should thus avoid completely (or at least minimise) execution of branches inside warps. Assigning a large number of warps to each SM (high occupancy) is beneficial as potentially long waiting times of some warp instructions can be conveniently hidden by executing instructions from other warps and selection of ready warps for execution does not introduce idle times into the overall execution achieving zero-overhead thread scheduling. 

\subsection{Performance Model}
\label{sec:model}

The main computation in Splotch occurs during the stages of {\it Rasterization} and {\it Rendering}.
The overall performance can be quantified as CPU and GPU processing times 
and time spent for performing data transfers among different memories.
% using the following expression:
%\begin{equation}\label{Ts}
%T_{TOT} = T_{cpu} + T_{pci} + T_{gpu} + T_{Mgpu},
%\end{equation}
%where $T_{TOT}$ denotes total times, 
%$T_{cpu}$ and $T_{gpu}$ represent the processing times of the CPU and GPU respectively, $T_{pci}$ are times needed for data transfers from the CPU to GPU and vice-versa via PCI-E, and finally $T_{Mgpu}$ are times required for data transfers between the global and shared memory on the GPU. 

\subsubsection{Data Transfers}
Let us denote by $S_{part}$ the number of bytes required for representing single particles (35~Bytes are used in the current implementation). Assuming an overall number of $N_{part}$ particles
%, the overall amount of bytes required to be transferred to the GPU is $N_{part} S_{part}$. Additionally the Splotch renderings calculated in the GPU are required to be transferred to the CPU. Assuming 
and square images (generalisation to rectangular images is straightforward) with $N_{pix}$ horizontal (or vertical) image resolutions, a number of $N_{part} S_{part} + 12 N_{pix}^2$~Bytes (R,~G,~B float values) has to be transferred. 
Since processor memory is in general larger than GPU memory,
datasets are split into chunks and are transferred to the GPU in a sequence one after the other. The overall data transfer time between CPU and GPU via PCI-E is:
\begin{equation}\label{pci}
T_{pci} =  N_{chunks} \tau_{pci} + {N_{part} S_{part} + 12 N_{pix}^2 \over 
\mu_{pci}},
\end{equation}
where $\tau_{pci}$, $\mu_{pci}$ represent transfer time latency and bus bandwidth and $N_{chunks}$ is the number of copy stages. Equation $\eqref{pci}$ shows that performance depends linearly not only on the number of particles but also on the image resolution. Assuming very large-scale datasets ($N_{part} >> N_{pix}^2$) the contribution of $N_{pix}^2$ to $T_{pci}$ becomes negligible. 
%(as an example see the test case described in Sect.~\ref{sec:results} in which $N_{part} \sim 10^8$ and $N_{pix}^2 \sim 10^6$).
Splitting datasets into chunks does not cause any meaningful overheads if the copy latency is small, that is:
\begin{equation}
N_{chunks} < {N_{part} S_{part}\over \tau_{pci}\mu_{pci}}.
\end{equation}
As an example, if $N_{part}$ is $O(10^8)$, then $N_{chunks}$ can be up to $10^5$ before
generating any overheads. 

\subsubsection{GPU Computation}
The time required for processing particles on the GPU can be estimated by:
\begin{equation}
T_{gpu} = N_{op}/\nu_{GPU}
\end{equation}
where $\nu_{GPU}$ is the GPU's flops/sec rate, and $N_{op}$ is the expected overall number of operations:
\begin{equation}\label{ops}
N_{op} = N_{part}(\alpha + \beta R_0^2) + f_{GPU},
\end{equation}
with $\alpha$ and $\beta$ representing the number of operations per particle in the \textit{Rasterization} and \textit{Rendering} kernels respectively.
The function $f_{GPU}$ accounts for GPU specific kernels; these are necessary for adapting the original Splotch code to the GPU architecture and optimizing performance. Details of these kernels are discussed in Sect.~\ref{sec:implementation} and Sect.~\ref{sec:results}.
Finally the parameter $R_0$ encapsulates an expectation value (in terms of number of pixels) for the radius of particles projected on Splotch renderings $R_0 = <r(p)>$,

\begin{equation}\label{radius}
r(p) = A(p){\chi \sigma_p\over S_{box}} N_{pix},
\end{equation} 
where $A(p)$ is the transformation to screen coordinates, $\chi$ and $\sigma_p$ are defined by equation \eqref{kernel} and $S_{box}$ represents 
normalisation factor measuring the size of the bounding box containing all the particles.
The radius of individual particles depends both on their intrinsic properties and on camera settings. %Although a particle may exhibit a negligible size from one point of view, it may affect the entire Splotch rendering from another point of view. Also, particles close to each other can have different radius depending on their intrinsic characteristics.

\begin{figure}
\centering
\includegraphics[scale=0.1]{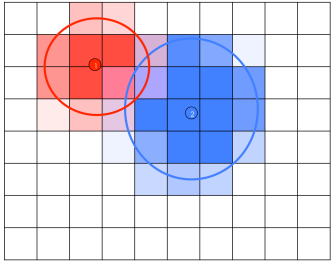}
\caption{Two particles with different radius influencing common screen pixels.}
\label{fig:particles}
\end{figure}

The dependency from $r(p)$ is the source of two major difficulties for GPU implementations. Firstly, two threads acting on different particles could try 
to update the same screen coordinates (see Fig.~\ref{fig:particles}) causing concurrent accesses to memory thus leading to erroneous results. 
Secondly, as different particles may affect different numbers of pixels, it is hard to achieve optimal load balancing among threads.  
Customised solutions must thus be adopted to circumvent the aforementioned problems while avoiding paying large performance penalties. 
%For instance, atomic updates (supported by CUDA) could be adopted for safely updating screen pixels, but this would result in serialization, strongly lowering performance. 

%To conclude this section we observe that the computing time depends linearly on the number of particles ($N_{part}$). This feature is important to be preserved in a GPU implementation of Splotch.

\subsubsection{Accessing Memory}
%Data transfers between global and shared memory can be expressed as the sum of the number of GPUs global to shared memory loads plus number of stores from shared to global memory:
%\begin{equation}
%N_{Mgpu} = (N_{load,p} + N_{store,p}) N_{part} + N_{store,pix} N_{pix}^2,
%\end{equation}
%where $N_{load,p}$ and $N_{store,p}$ are the number of loads and stores of individual particles, and $N_{store,pix}$ is the number of stores of rendered images (no loads are expected since images are created on the GPU). Furthermore, memory access overheads can be estimated as $N_{Ngpu} \nu_{mem}$; $N_{Ngpu}$ is the overall number of memory accesses and $\nu_{mem}$ represents memory frequency. The overall time required for moving data among memories is thus calculated by:
%\begin{equation}\label{tmgpu}
%\begin{split}
%T_{Mgpu} = {(N_{load,p} + N_{store,p}) N_{part} S_{part}
%+ 12 N_{store,pix} N_{pix}^2\over \mu_{gpu}} + \\
%+ N_{Ngpu} \nu_{mem} + g_{GPU},
%\end{split}
%\end{equation}
%where $\mu_{gpu}$ is the global memory bandwidth and $g_{GPU}$ the time spent on memory accesses by GPU specific functions. Equation \eqref{tmgpu} shows that 
Performance related to global/shared memory data transfers depends on the number of loads $N_{load,p}$ and stores $N_{store,p}$ of individual particles and on memory bandwidth.
An optimal solution would be to realise a single data transfer only
(i.e.\ $N_{load,p} = 1$ and $N_{store,p} = 0$) with each thread processing a different particle fully. Such an approach ({\it one-thread-per-particle}) could guarantee optimal exploitation of GPU architectures. However,
 this is not practical, the main reason being that frequent race conditions may arise as discussed already.
Only our {\it Rasterization} kernel is data parallel and can thus implement the {\it one-thread-per-particle} approach. 
%For this stage $N_{load,p} = 1$ and since processed particles must be copied back to the global memory to be rendered $N_{store,p} = 1$. 
The {\it Rendering} kernel is envisaged to require an additional data load, but no more store 
stages are necessary.
% (particles can be ``forgotten'' once their contribution is calculated).

The contribution to the overall time required for moving data among memories related to the image size is negligible for the cases relevant to this work, 
for which we generally have $N_{part} >>  N_{pix}^2$. Exceptions can be related to peculiar camera positions, for which the number of  active particles can become comparable to the number of pixels. 

Another critical term is the memory access latency. 
Since access to memory is slow compared to
the available bandwidth, 
%so a small number of accesses is critical for achieving high performance. This can be realised by standard caching strategies, that however are effective only if 
data coalesced access should be guaranteed (i.e.\ data are accessed by warps in one single memory transaction). In particular this happens when consecutive particle data moved to the GPU L2 cache can be reused efficiently by consecutive threads. 

\section{CUDA Implementation}
\label{sec:implementation}

Once particles are loaded into the GPU's global memory, the {\it Rasterization} kernel starts processing them. We follow an efficient {\it one-thread-per-particle} approach, so that the entire computation on individual particles (e.g.~data normalisation\footnote{Tipically min/max normalisation values are specified by the user, otherwise they are pre-computed on the host.}, geometric transformation and coloring) is carried out by single threads - no interaction with any other processes is necessary. 
Subsequently the {\it Rendering} phase takes over. As explained previously (see Sect.~\ref{sec:model}) we cannot adopt a {\it one-thread-per-particle} approach and memory usage must be managed carefully. To handle particles in an efficient way we observe that depending upon their radius ($r(p)$) as defined by equation \eqref{radius} the rendering process can be performed differently.

\begin{figure}
\centering
\includegraphics[scale=0.14]{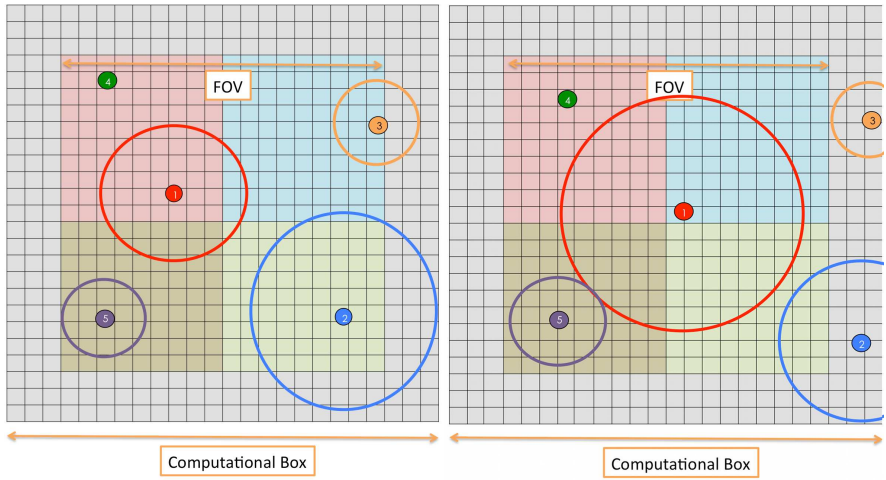}
\caption{A scene with five particles from two different points of view. Square tiles are
represented with different colors and boundary width is $r_0=3$ pixels. {\em Left}: All particles contribute to the image (inside the field of view), particles 1, 3 and 5
are medium, particle 4 is small. Particle~2, is classified as large since its radius exceeds the boundary width. {\em Right}: The camera has now moved towards particle~1. The relevant particle radii change due to the new point
of view. Classification of particles 4 and 5 does not change, while particle~3
becomes inactive (completely falling outside the field of view). Note that although particle~2 is outside the field of view it is designated as active as it still affects pixels.
Finally particle~1 becomes large.
}
\label{fig:fov}
\end{figure}

Considering particles with a small radius, these typically influence single pixels on the rendered images, so we can process them again by deploying a {\it one-thread-per-particle} approach, but carefully managing possible race conditions (see Sect.~\ref{sec:smallparticles}). On the other hand, particles with very large radius affect sizeable areas of the rendered images, so these are not suitable for efficient GPU processing (see Sect.~\ref{sec:largeparticles}). Finally, particles with radius within a suitably defined range influence only small fractions of the rendered images, so we process these by using an efficient tiling strategy (see Sect.~\ref{sec:mediumparticles}). We classify particles based on the size of their radius as follows:

\begin{itemize}
\item
{\it Small (S)} : $r(p) \le r_{sml}$,
\item
{\it Medium (M)} : $r_{sml} < r(p) \le r_0$,
\item
{\it Large (L)} : $r(p) > r_0$.
\end{itemize}
where $r_{sml} = 0.5$ so that small particles fall inside single pixels, and 
$r_0$ is set according to an image tiling scheme (see Sect.~\ref{sec:mediumparticles} for details) which divides the rendered images
into a number of tiles as follows:  
\begin{equation}\label{tiles}
N_{tiles} = {N_{pix}^2\over t_x \times t_y},
\end{equation}
where $t_x$ and $t_y$ are parameters defining the sizes of tile sides in pixels.
An example of particle classification from different points of view is shown in Fig.~\ref{fig:fov}. Each particle is labeled (in the {\it Rasterization} kernel) with a {\it tile index} $i(p)$ which is calculated as follows:
\begin{itemize}
\item
$i(p) = -2$, {\it non-active} particles (i.e.\ outside the field of view);
\item
$i(p) = N_{tiles}$, {\it small} particles;
\item
$0 \le i(p) < N_{tiles}$, {\it medium} particles, whose center falls within the $i(p)$-th tile;
\item
$i(p) = -1$, {\it large} particles.
\end{itemize}

\begin{figure}
\centering
\includegraphics[scale=0.6]{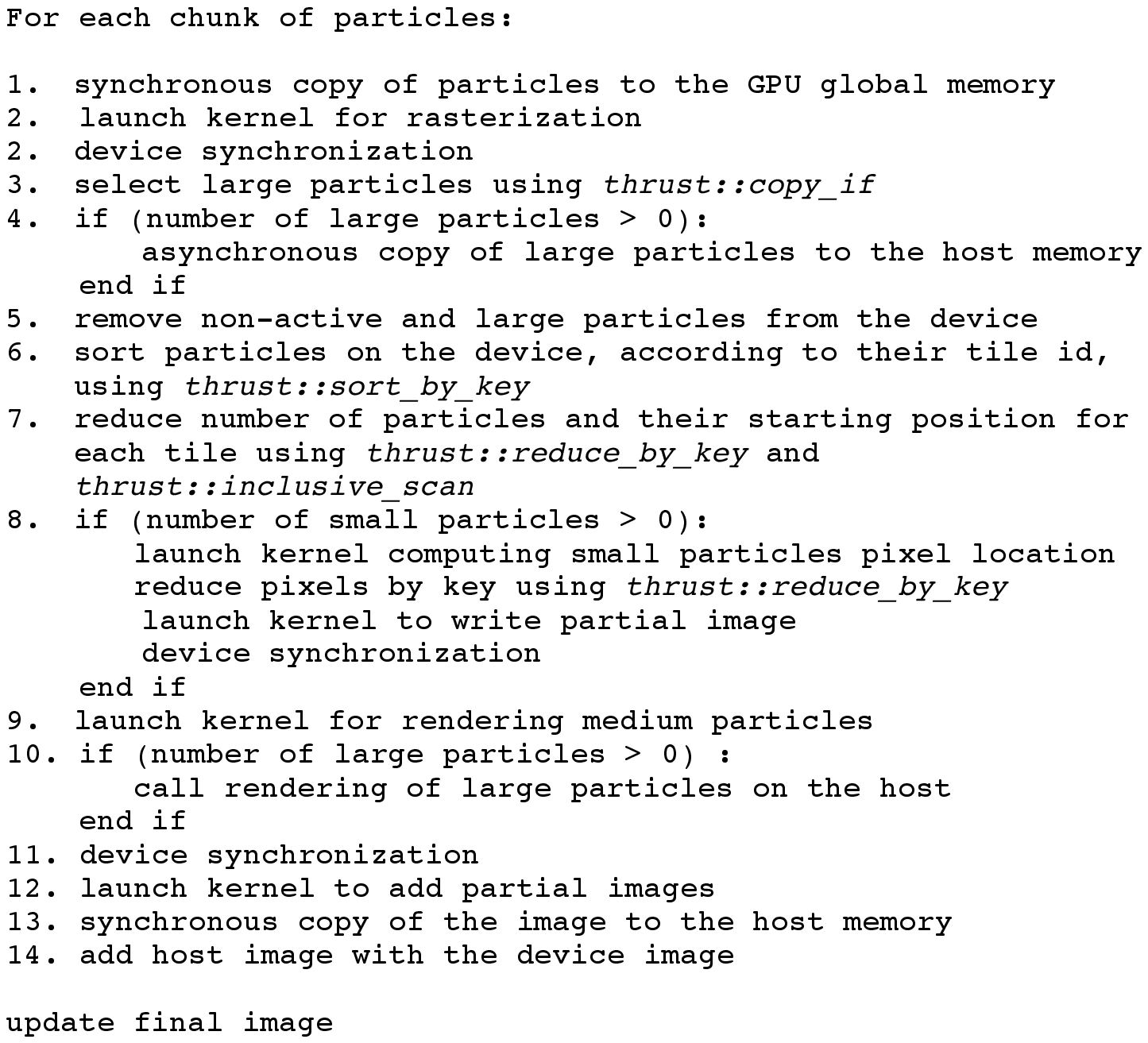}
\caption{Pseudo-code.
}
\label{fig:pseudo-code}
\end{figure}

Subsequent to classification we transfer asynchronously all large particles back to the host and remove them from the device memory (at the same time all inactive particles are also removed). At this point, the remaining particles have to be sorted by the $i(p)$ key and the number of particles with the same index has to be calculated.
This sorting operation is important as it allows to manage a particles array on the device and it is  
necessary for the efficient execution of subsequent operations, e.g. reduction by key and prefix sums.
As sorting is intrinsically an expensive operation, we use an efficient implementation provided by the Thrust library \cite{thrusturl}. Thrust is a C++ template library for CUDA which mimics the Standard Template Library and provides optimised functions for managing very large data arrays. For our purposes, Thrust implements a highly optimised Radix Sort algorithm for primitive types (e.g.\ chars, ints, floats, and doubles). Also, functions are provided for reduction by key and prefix sums for which their performance increases as the input data arrays become larger. As these functions scale linearly with data sizes (sorting scales by $kN_{part}$, where $k$ is the number of significant key bits \cite{RadixSort}), our GPU implementation preserves the original Splotch linear dependency on the number of particles. The sections below describe rendering details for individual classes of particles and Fig.~\ref{fig:pseudo-code} contains a description of the code structure.

\subsection{Small Particles}
\label{sec:smallparticles}
To process small particles only their location within rendered images has to be calculated. This can be efficiently carried out by a simple CUDA kernel. However, since different threads can affect common pixels in the rendered images, their contribution (fragments) cannot be directly accumulated. Thus, first we allocate an index buffer in the device memory to store particle positions within the image (pixel index). Second, the rendering is updated by reducing by key (key being the pixel index) all computed fragments by using Thrust functions and adding them to the image in another kernel.

\subsection{Medium Particles}
\label{sec:mediumparticles}
Our \textit{tiling scheme} for processing medium particles on the device consists of assigning particles related to the same image tile (as defined by \eqref{tiles}) to a block of CUDA threads and exploiting  the shared memory and thread synchronization within the block to store and compose image tiles. The local size of the tile is defined so that each particle belonging to it, is entirely contained within it. This is achieved by adding to the $body$ of the tile of $t_x \times t_y$ pixels a $boundary$ of $r_0$ pixels around it, requiring that $r_0 \le \min\{t_x, t_y\}$ in order to avoid the boundary to overlap more than one neighbouring tile. We will refer to this extended tile as a \textit{Btile}.

Particles are accessed in chunks of $n_p$ elements. Each chunk is accessed and stored in the shared memory by a single read operation performed simultaneously by the threads of the block. Then, the particles of the chunk are rendered sequentially, each with a single parallel operation so that each pixel affected by a particle is processed by a different thread of the block (although the pixel number processed by each thread may change as the particle varies, it always remains in the Btile). The block size is equal to the number of pixels of the largest medium particle allowed (i.e.~$4r_0^2$), in order to have enough threads in the block for rendering each pixel influenced by the current particle. 
This solution avoids race conditions when composing image tiles, since each
thread of the same block accesses different pixels. The workload per particle of each
thread is well balanced as each thread processes either one or no pixels at all for each particle, independently of the size of individual particles.

When all particles of the block are rendered, the contribution of the Btile is incorporated to the image stored in global memory. Specific care has been taken to avoid race conditions due to contributions coming
from overlapping regions (boundaries). Since CUDA blocks are order independent, three more copies of the image in the global memory are required to store corners, rows and columns of the boundary respectively. The final rendering is obtained through a second kernel compositing these copies with the partial image produced by the small particles.  

\subsection{Large Particles}
\label{sec:largeparticles}
Large particles are the most challenging to process as their large
radius prevents the usage both of a tiles based solution and of a fragment buffer.
In the first case, tiles would be too large to be stored in the shared memory
and, in any case, their large overlap could lead to strong overheads in the composition of the
final rendering. Using a fragment buffer instead would require too much memory since
for each particle a large number of fragments could be generated, possibly
exceeding in size the overall available memory.

The implemented solution consists in copying these particles back to the CPU and performing their rendering
using the serial version of Splotch. This is possible thanks to CUDA asynchronous
operations, which allow to copy data from the device to the host (and vice-versa)
whilst the calculation on the GPU is on-going. Once the particles are back to the CPU,
their processing can be performed concurrently to that of the GPU.  
This way, we manage to exploit both the host and the device simultaneously.
This solution is effective as long as the number of large particles
is much smaller than the sum of small and medium particles. As a worst case scenario (although highly unlikely in practice), if all particles are large the performance would degenerate to that of serial Splotch.

Once all calculations are completed, two partial images (one processed by the GPU the other by the CPU) are composed to generate the final rendering. Such operation is performed by the CPU, once the GPU image has been transferred in a single copy operation. 

\subsection{Remarks}
Note that for both small and medium particles the main drawback of the proposed solutions
is represented by the overhead associated with the sorting and reduction operations. Specifically the sorting is intrinsically time-consuming requiring intensive usage of memory.
The performance of rendering medium particles is mainly influenced
by the size of shared memory which limits the number of resident blocks
per multiprocessor during the execution of the kernel, thus reducing device occupancy. 
A further issue is related to the unbalanced work load among blocks; the number of particles within each tile can vary significantly. All these aspects are discussed and quantitatively analyzed in Sect.~\ref{sec:results}.

\section{Results}
\label{sec:results}

An N-body-SPH simulation performed using Gadget \cite{gadgeturl} was used for the tests we discuss in this section. The simulation is based on about 400 million particles consisting of 200 million {\it dark matter} particles, 200 million baryonic matter particles (hereafter referred to as {\it gas}) and around 10 million {\it star} particles. Particles possess a number of physical quantities, e.g.\ vectors capturing spatial coordinates and velocities, or scalars representing smoothing length. Additionally gas particles are associated with {\it temperature} and {\it mass density}. Also, {\it spectral type} and {\it age} are properties of star particles. Such quantities can be exploited in highly-detailed renderings, e.g.\ by modulating colors and intensities appropriately. 

We used GCC 4.8.1 and CUDA~5.5 for compilation. Execution was performed on a computing node containing a dual-socket 8-core Intel Xeon E5-2670 processor with 32~GB shared memory and one NVIDIA Tesla~K20X GPUs with 6~GB of GDDR5 memory, 250~GB/sec main memory bandwidth and peak single precision floating point performance of 3.95~TFlops/sec. Although such a large-memory computing node is necessary for handling our test simulation ($\approx 7.5$~GB) no GPU issues arise as particles are off-loaded in chunks of suitable size. 

\subsection{Tuning}
\label{sec:gpuperf}
An important aspect, for GPU performance tuning, is ensuring high {\it occupancy} to avoid cores remaining idle during kernel execution. Occupancy depends upon the number of threads per block, the number of registers and the size of shared memory used in kernels - these define the number of resident CUDA blocks (blocks/SM) during execution. We have empirically estimated a trade-off among these parameters for the rendering kernel of our implementation. This kernel may potentially use significant shared memory due to the fact that Btiles together with a number of particles need to be stored. As the boundary width of Btiles is critical for realizing particle classification and determining sizes of CUDA blocks, we suggest to set this no larger than 8 pixels due to the typically small size of the shared memory and focus instead on optimizing the relevant number of particles ($n_p$) and the size of tiles (for simplicity we assume square tiles $t_x^2$ only) respectively.

A number of possible $n_p$ and $t_x$ values are illustrated in Table~\ref{tab:tuning}. We consider a situation in which the majority of particles is non-large so that they are processed by the GPU. This way we obtain meaningful rendering times for our optimization as the host computation is indeed overlapped by the device computation completely. As expected, storing Btiles together with a fixed number of particles in the shared memory (for optimizing memory access) restricts full GPU occupancy. The maximum real occupancy ($75\%$) and the best CUDA times are obtained for tiles with $t_x = 8$ and chunks of  $n_p = 16$ particles\footnote{On Fermi architecture, the maximum occupancy obtained is $66.7\%$ for tiles with $t_x = 10$ or $t_x = 12$ and chunks of particles with $n_p = 32$ or $n_p=64$. Taking into account total CUDA times, optimal performance is achieved for $n_p=64$ and $t_x = 12$.}, so we employ these values for testing in the next sections.

\begin{table}
\label{tab:tuning}
\begin{center}
\begin{tabular}{llrlll}
\hline \noalign{\smallskip}
$n_p$ & $t_x$ & Shared Mem. & Kernel & Rendering & Total CUDA \\
            & & (Bytes) & Occupancy & Times (sec.) & Times (sec.) \\
\noalign{\smallskip} \hline \noalign{\smallskip}
64   & 12 & 11,864 & 0.5 & 5.268 & 9.228 \\
      & 10 & 10,928 & 0.5 & 5.087  & 9.045 \\
      & 8 & 9,728 & 0.625 & 5.334  & 9.301 \\
32   & 12 & 10,456 & 0.5 & 5.379 & 9.344 \\
      & 10 & 9,520 & 0.625 & 4.225  & 8.185 \\
      & 8 & 8,320 & 0.625 & 4.410  & 8.379 \\
16  & 12 & 9,752 & 0.5 & 5.588 & 9.538 \\
      & 10 & 8,816 & 0.625 & 4.399 & 8.363 \\
      & 8 & 7,616 & 0.75 & 3.889 & 7.873 \\
\noalign{\smallskip} \hline
\end{tabular}
\caption{The occupancy and rendering times obtained for different values of $n_p$ (number of particles stored in the shared memory) and $t_x$ (tile sizes). Square tiles with boundary width 8 are considered while the resolution of Splotch rendered images is $1024^{2}.$}
\end{center}
\end{table}

Therefore, for typical image sizes ($N_{pix} > 10^3$), considerations on occupancy lead to choosing  very small tile sizes. Also load balancing benefits of this choice, since as tile size decreases, the number of particles per tile decreases as well, and the workload results to be evenly distributed.
The main drawback of this choice is that for small tiles, a larger number of particles are classified as large, hence processed on the CPU. We will show in the next sections the overall quantitative consequences of this setup.  

\subsection{Scalability}
\label{sec:scalability}
\begin{figure}
\centering
\includegraphics[scale=0.5]{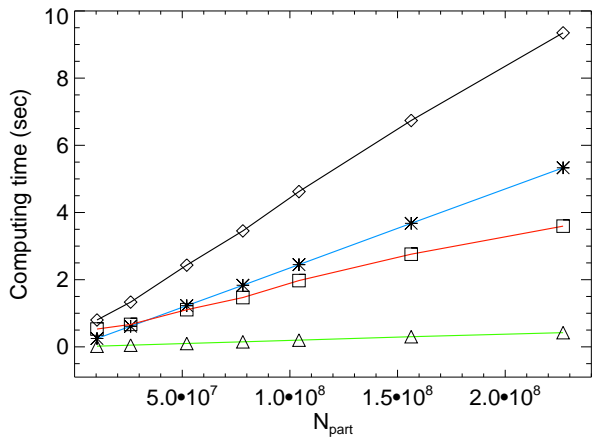}
\caption{Scalability of Splotch kernels on the GPU in relation to dataset sizes. 
The lines are colored as follows: a) black (diamonds) showing total computing times, b) blue (crosses) showing rendering times, c) red (squares) representing overhead times and finally d) green (triangles) illustrating rasterization times.}
\label{fig:scalability}
\end{figure}

To analyze scalability in terms of sizes of datasets we fixed the size of rendered images and camera settings so that all particles are classified as non-large and are thus processed by the GPU entirely. 
Figure~\ref{fig:scalability} illustrates the obtained timings when progressively increasing the overall number of particles
from about $10^7$ to roughly $2.2\times10^8$. Computing times scale linearly confirming that this feature of the original CPU Splotch is preserved. 

Linear scalability with $N_{part}$ is measured for both the main contributors to processing time: the rendering time and the overheads, related to all those parts specific to the GPU code refactoring (e.g.\ times for host-device copy or distribution of particles - see Sect.~\ref{sec:overhead}).
Rasterization scales linearly as well, and represents always a minor contribution to the overall computing time.   

\begin{figure}
\centering
\includegraphics[scale=0.5]{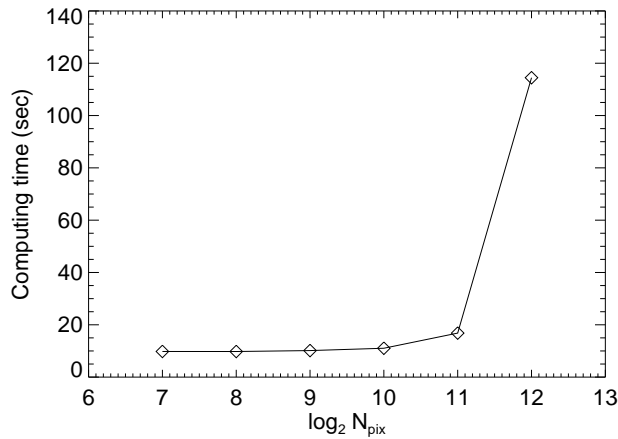}
\caption{Splotch's scalability on the GPU in relation to image side $N_{pix}$ (camera settings are fixed).}
\label{fig:pixels}
\end{figure}

Overall performance in relation to image sizes $N_{pix}^2$ is illustrated in Fig.~\ref{fig:pixels}. Keeping the underlying dataset size and camera settings fixed, we obtained computing times by progressively increasing
$N_{pix}$ from ${2}^{7}$ to ${2}^{12}$. According to \eqref{ops} and \eqref{radius} each particle contributes to a number of pixels that is proportional to $N_{pix}^{2}$. As long as image sizes are small relative to dataset sizes (for our particular configuration this translates to $N_{pix} \le 2^{11}$), overall computing times are constant. For large image sizes their resolution may affect significantly the classification of particles potentially increasing overall computing times.

\subsection{Speed-up}
\label{sec:speed-up}

\begin{figure}
\centering
\includegraphics[scale=0.22]{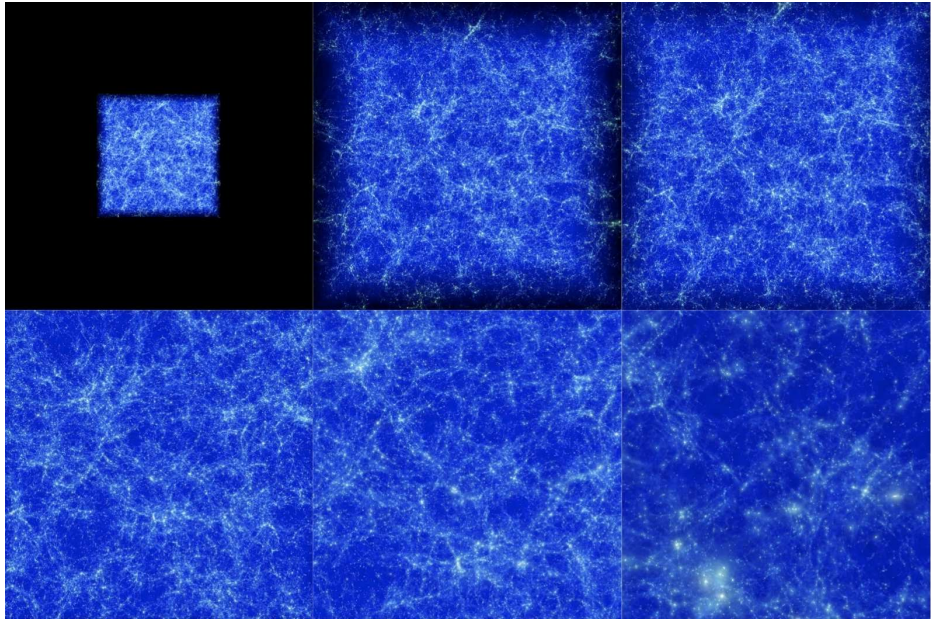}
\caption{Our reference particle simulation as rendered by Splotch at a number of different camera positions mimicing a zoom-in operation. Starting from very far (top-left, Test 1) and reaching progressively very close to the center of the simulation (bottom-left, Test 6 - the image of Test 7 is omitted). The mass distribution of gas particles is shown.}
\label{fig:panorama}
\end{figure}

In the next series of tests, we discuss the speed-up achieved with respect to the original Splotch. 
The dependency of the computing time from the radius, defined by equation \eqref{radius}, has been analysed maintaining a fixed image resolution at ($N_{pix} = 2^{11}$) and progressively changing the camera position 
so as to resemble a typical user interaction scenario, namely a zoom in operation. 
The starting camera position ensures that the entire box of the simulation is rendered (Test~1). 
Next we ensure that the simulation box just about fits into the rendered images (Test~2). 
A few other camera positions have also been set so as to reach closer and closer towards the 
centre of the simulation (Tests~3 to~7). Figure~\ref{fig:panorama} illustrates rendered images generated for the different camera positions.

\begin{table}
\label{tab:bin}
\begin{center}
\begin{tabular}{lrrr}
\hline\noalign{\smallskip}
Test ID & $R_0$ (pixels) & Active Particles & Time (sec.) \\
\noalign{\smallskip} \hline \noalign{\smallskip}
Test 1  & 0.30   & 226,894,837  & 10.14 \\
Test 2  & 0.62   & 225,972,201  & 10.65 \\
Test 3  & 0.72   & 212,746,328  & 10.36 \\
Test 4  & 1.40   & 153,647,633  & 8.83 \\
Test 5  & 3.90   & 64,756,141   & 10.04 \\
Test 6  & 8.99   & 13,686,588   & 11.35 \\
Test 7  & 13.66  & 4,222,214    & 9.07 \\
\noalign{\smallskip} \hline
\end{tabular}
\end{center}
\caption{Illustration of average particle radius (column 2) together with number of active particles (column~3) and computing times (column~4) for each test performed (test ID is displayed in column~1).}
\end{table}

\begin{figure}
\centering
\includegraphics[scale=0.5]{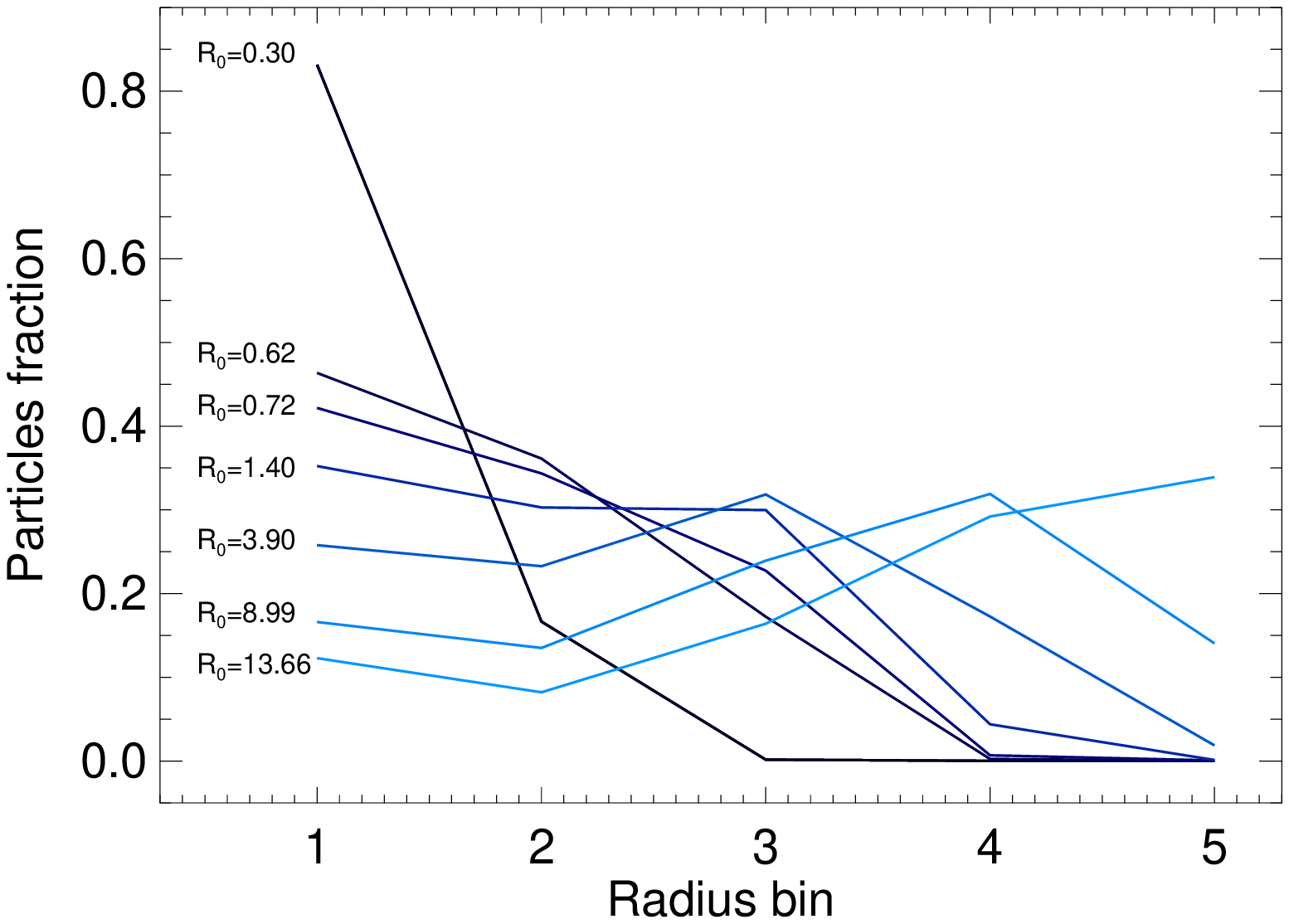}
\caption{Normalised percentage distribution of the particles radii in the seven tests (dark blue line refers to Test~1, light blue to Test~7). Radius bins in the x-axis corresponds to: $r\le 2$ (bin~1), $2<r\le 4$ (bin~2), $4<r\le 8$ (bin~3), $8<r\le 16$ (bin~4), $r>16$ (bin~5). 
}
\label{fig:radii}
\end{figure}

We observe that approaching the center of the simulation results in larger radii 
as expressed by the average $R_0=<r(p)>$ radius and illustrated by Fig.~\ref{fig:radii}. 
Table~\ref{tab:bin} presents the estimated values of $R_0$, together with the number of active particles and computing times. As we approach the centre of the simulation fewer particles are active, an increasingly larger number being outside the scene - this leads to a reduction in computing times. At the same time, however, particles are closer to the camera and their radius increases contributing to a progressively larger number of screen pixels,
thus increasing computing times with the square of their radius. 
These trends tend to compensate maintaining on average~9 to~11 seconds overall computing times.

\begin{figure}
\centering
\includegraphics[scale=0.5]{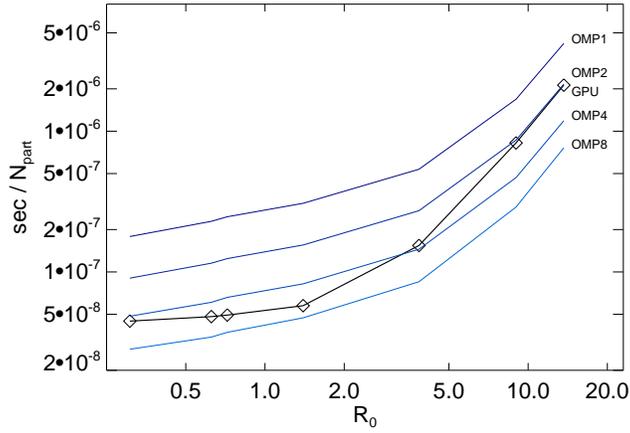}
\caption{GPU and CPU comparison in relation to the particles' radius. The black curve represents GPU computing times. Blue curves are CPU times ranging from 8 cores (light blue, OMP8) to 1 core (dark blue, OMP1).}
\label{fig:gpucpu}
\end{figure}

CPU Splotch timings were obtained adopting the OpenMP paradigm on 1, 2, 4 and 8 cores respectively. OpenMP was used in order to exploit the shared memory capability of the computing node. It provides a slightly better intra-node parallel performance with respect to MPI. The results are illustrated in Fig.~\ref{fig:gpucpu} as a function of computing times per particle, since the number of processed particles depends on camera settings.

Note that the GPU curve exhibits a bi-modal behaviour depending upon the radius. For small values, computing time increases slowly, since the overall calculation is dominated by $N_{part}$. For values larger than unity the dependency from the square of the radius dominates and computing time rises rapidly. The comparison between the GPU and the multicore performance show that at small radii ($R_0 < 2$) the speed-up gained by the GPU implementation can be larger than a factor of~5 compared to a single CPU's core. Considering the OpenMP implementation of Splotch we observe that around unit radii, GPU performance is close to that obtained from~8 cores. At large radii ($R_0 > 2$) the performance of CUDA deteriorates naturally as the majority of particles is processed by the CPU. Nevertheless assuming a worst case scenario GPU Splotch is always faster than two cores execution. 

\begin{figure}
\centering
\includegraphics[scale=0.5]{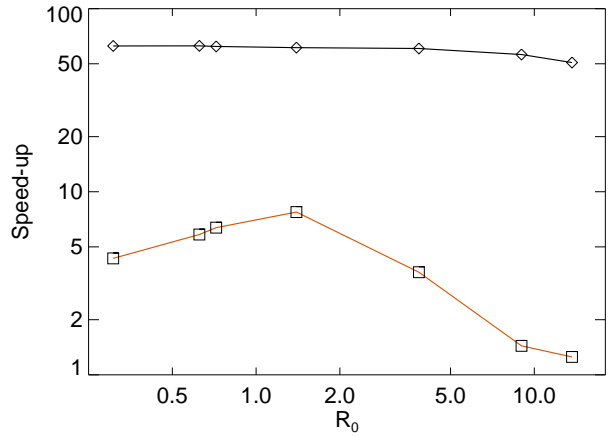}
\caption{
Observed speed-ups for rendering (red curve, squares) and rasterization (black curve, diamonds) kernels defined as ratios among computing times required for execution by the GPU and CPU respectively.
}
\label{fig:speedup}
\end{figure}

We conclude this section with a discussion on the performance achieved by all GPU kernels implemented. 
We compare computing times required for rasterization and rendering on the 
GPU and CPU respectively (see Fig.~\ref{fig:speedup}). 
As regards the rendering kernel, a speed-up of around~8 is obtained for average radii around unity. For smaller radii the speed-up
tends to decrease due to the lower occupancy of the rendering kernel processing medium particles.
In fact, from Fig.~\ref{fig:radii}, one can see that for $R_0<1.4$ most of the medium particles have a radius lower than the threshold~$r_0=8$ defining the size of CUDA blocks, therefore several threads in each block do not process any pixel.
As the radius grows, performance decreases due to the increasing number of large particles.
Regarding the rasterization kernel due to the {\it one-thread-per-particle} approach it is strongly accelerated with observed speed-up ranging from~50 to~63, this fluctuation depending on the number of active particles (which significantly reduces at large radii) leading to a slight drop of gain. 
%less computation performed, therefore less speed-up with respect to the CPU.

\subsection{Overheads}
\label{sec:overhead}
\begin{figure}
\centering
\includegraphics[scale=0.5]{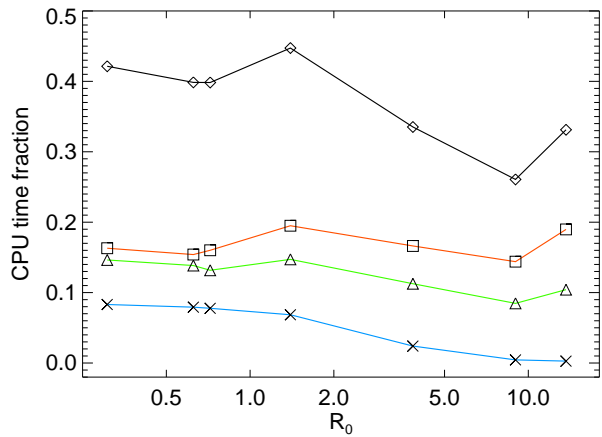}
\caption{Overhead of the various components implemented for the Splotch's
GPU refactoring. The ratio between the time spent in each component and the total processing 
time is illustrated. The black line (diamonds) shows total overhead, the green line (triangles) shows the Select time overhead, while the red line (squares) represents the OffLoad time, and finally the blue line (crosses) shows the Sort time.}
\label{fig:over}
\end{figure}

We now discuss the GPU overheads, i.e.\ times required by functions specific to our GPU implementation. To this extent a number of operations are necessary, e.g.\ for offloading particles to GPU (OffLoad), sorting and reducing for preparing Btiles and optimizing memory displacement (Sort), removing non-active particles, but also packing and copying back particles that have to be rendered by the CPU (Select) and performing partial images composition. Fig.~\ref{fig:over} presents the overhead of different GPU specific functions as times spent by the function divided by overall computing times. The black curve shows total overheads which can be up to~45\% and particularly significant for small radii. As most of the particles are active the sort and reduce
functions (blue line with crosses) are computationally demanding. For large radii this overhead is negligible as most of the particles are inactive. 
The red curve represents offload times, which are comparable throughout the tests (as all particles are moved to the GPU irrespectively of camera position), and is between 14 and 20\% of the total processing time. This contribution varies according to the total processing time only, as the offload time is independent of the particles radius.
The Select times (green line) are slightly lower than OffLoad. This difference tends to increase at larger radii,  when more particles are moved back to the CPU, hence coalesced memory access can be effectively exploited. 
%when less particles contribute to the image rendering. 
Sorting (blue line) is always below the 10\% of the processing time and its contribution tends to strongly decrease with decreasing the 
number of active particles.
Finally, image composition (not shown in the figure), and any other CUDA set up times, gives always a negligible contributions, e.g.\ creating rendered images accounts for approximately~0.1\% of GPU overheads. This is due to the efficient design of our implementation taking full advantage of the multi-thread architecture of GPUs.

\section{Conclusions and Future Work}
\label{sec:conclusions}
The CUDA implementation of Splotch exploits GPUs in modern HPC infrastructures for visualizing large data volumes such as those produced by experiments/observations and by computer simulations. Current trends in HPC envisage the widespread adoption of accelerators in the coming years to substantially increase computing performance while maintaining power consumption reasonably low. However, accelerators, and in particular GPUs, demand a certain effort to be efficiently exploited, as they support new programming models typically requiring entire codes, or part of them, to be redesigned and reimplemented.

Splotch proved to be a challenging algorithm to adapt for GPU execution. The main computational kernel poses serious difficulties to the GPU's programming model which strongly favours highly parallelised approaches. We re-designed Splotch by introducing a number of specific kernels which may involve overheads of up to~45\% of overall computing times. Nevertheless the original sequential 
version on Intel Sandy Bridge processor is outperformed on Kepler architecture by a factor of~5, giving performance comparable up to 8 CPU cores when the OpenMP multithread parallelised version is considered for comparison. An important aspect is retaining linear scalability on the number of particles processed.

Further optimizations could be obtained by adopting CUDA streams, overlapping computations and data transfers. However this is currently prevented by the rendering of large particles on the CPU and the Thrust library which provides no support for such operational scenarios.

The achieved performance gains allow exploitation of hybrid computing nodes nowadays increasingly common in HPC infrastructures. However to fully exploit such architectures further work is required towards concurrent usage of multiple GPUs, multiple cores and multiple nodes, and exploitation of the OpenMP and MPI Splotch capability. This work is expected to be challenging especially in the context of finding an optimal balance in the workload of individual system components. Additionally novel solutions provided by Kepler architecture~\cite{GK110}, such as dynamic parallelism and Hyper-Q usage of multicore systems, should be explored. This last feature is particularly relevant for our purposes since it enables multiple CPU cores to simultaneously utilise CUDA cores on a single GPU. We envisage this to be particularly effective when off-loaded data can be split in small chunks and progressively copied by different MPI tasks to the GPU memory (see~\cite{GPUTech}). 

The vision is to further develop Splotch as an effective approach for large data visualization and discovery, capable of optimally exploiting any underlying HPC devices irrespectively of their specific architectures. In this respect, we are also experimenting with the novel INTEL Xeon Phi~\cite{mic}, which represents a further interesting solution based on accelerators. Results will be presented in a forthcoming article~\cite{mic-splotch}.
Furthermore, we are also developing specialised utilities such as a lightweight interactive rendering engine mimicking Splotch's imagery for rapid inspections of large-scale astrophysical datasets.

\bibliography{master}	

\end{document}